\documentclass[twocolumn,showpacs,preprintnumbers,amsmath,amssymb]{revtex4}

\usepackage{graphicx}
\usepackage{dcolumn}
\usepackage{bm}

\begin{document}

\title{Low-Energy Photodisintegration of the Deuteron and Big-Bang 
Nucleosynthesis}

\author{W. Tornow$^{a,b}$, N.G. Czakon$^b$, C.R. Howell$^{a,b}$, 
A. Hutcheson$^{a,b}$, J.H. Kelley$^{c,b}$, V.N. Litvinenko$^{a,d}$, 
S. Mikhailov$^{a,d}$, I.V. Pinayev$^{a,d}$, G.J. Weisel$^e$, H. Wita{\l}a$^f$}

\affiliation{$^a$Physics Department, Duke University, Durham, NC 27708, USA \\
$^b$Triangle Universities Nuclear Laboratory, Durham, NC 27708, USA\\
$^c$Physics Department, North Carolina State University, Raleigh, NC 27695, USA\\
$^d$Duke University Free-Electron Laser Laboratory, Durham, NC 27708, USA\\
$^e$Physics Department, Penn State Altoona, Altoona, PA 16601, USA\\
$^f$Institute of Physics, Jagiellonian University, PL-30059 Cracow, Poland
}%

\date{\today}

\begin{abstract}
The photon analyzing power for the photodisintegration of the deuteron was 
measured for seven gamma-ray energies between 2.39 and 4.05~MeV using the 
linearly polarized gamma-ray beam of the High-Intensity Gamma-ray Source at
the Duke Free-Electron Laser Laboratory.  The data provide a stringent test 
of theoretical calculations for the inverse reaction, the neutron-proton 
radiative capture reaction at energies important for Big-Bang Nucleosynthesis.
Our data are in excellent agreement with potential model and effective field
theory calculations.  Therefore, the uncertainty in the baryon density
$\Omega_Bh^2$ obtained from Big-Bang Nucleosynthesis can be reduced at
least by 20\%. 
\end{abstract}


\maketitle

Big-Bang Nucleosynthesis (BBN) is an observational cornerstone of the hot 
Big-Bang (BB) cosmology.  According to \cite{Bur99} the neutron($n$)-proton($p$) 
capture reaction $p(n,\gamma)d$ with a deuteron ($d$) 
and a 2.225 MeV $\gamma$ ray in the exit channel is of special interest, 
because the BB abundance of deuterium provides direct information on the baryon
density in the early universe at times between about 0.01 and 200 
seconds after the BB.  Knowing accurately the $n$-$p$ capture cross section 
in the energy range from 25 to 200 keV in the center-of-mass (c.m.) system and 
using the experimental value for the primeval deuterium number density 
(D/H)$_p$ \cite{Bur98a,Bur98b}, would allow for an accurate determination of 
the baryon density $\Omega_Bh^2$ ($h$ is the Hubble constant in units of 
100 km/s/Mpc).  
From $\Omega_Bh^2$ one can predict the abundances of
the three light elements $^3$He, $^4$He, and $^7$Li.  According to 
\cite{Bur99},
the 10\% uncertainty in the deuterium-inferred baryon density $\Omega_Bh^2
= 0.019 \pm 0.002$ comes in almost
equal parts from the (D/H) measurements and theoretical uncertainties in 
predicting the deuterium abundance.  For the latter, the knowledge of the 
$n$-$p$ capture cross section is of crucial importance.  Unfortunately, there
is a near-complete lack of data at energies relevant to BBN.  Aside from 
thermal energies, data exist only at $n$-$p$ c.m.\ energies of 275 keV and 
above.  As a consequence, the ENDF-B/VI \cite{Hal97} evaluation has been used 
\cite{Bur99} in the BBN energy range.  This evaluation is normalized to the 
high-precision thermal $n$-$p$ capture cross-section measurements. The
5\% uncertainty that is assigned in this approach contributes a significant 
fraction to the 
uncertainty in the baryon density and consequently in the abundances of the 
light elements produced in BBN.

Very recently, with the precision results from WMAP (Wilkinson Microwave 
Anisotropy Probe) for the Cosmic Microwave Background (CMB) and its 
anisotropies an independent and even more accurate result became available:
$\Omega_Bh^2 = 0.0224 \pm 0.0009$ \cite{Ben,Spe}.  The comparison of the baryon
density predictions from BBN and the CMB is a fundamental test of BB cosmology
\cite{Cyb}.  Any deviation points to either unknown systematics or the need for
new physics.  Therefore, it is of crucial importance to reduce the uncertainty
in $\Omega_Bh^2$ obtained from BBN.  As stated above, 50\% of the uncertainty
is due to the uncertainty in the $n$-$p$ capture cross section in the energy
range of interest.

Recently, effective field theory approaches \cite{Che99,Rup00} have provided 
accurate results for the time-reversed reaction $\gamma$-$d \rightarrow
n$-$p$ from threshold (2.225 MeV) to about 10 MeV incident $\gamma$-ray
energy.  The work described in this Letter was motivated by these new 
theoretical results
and also earlier nucleon-nucleon potential model based calculations \cite{Are}
in the $\gamma$-ray energy range important to BBN ($E_{\gamma}$ = 
2.25--2.43~MeV).  Aside from the $\gamma$-$d$ cross section, and the 
$n$-$p$ capture cross section inferred via ``detailed balance,'' these 
calculations
predict results for other observables as well which are related to the cross
section, but are in principle experimentally easier to measure with high
accuracy than the $n$-$p$ capture cross section itself.  The aim of this work
is to provide an alternative method of determining the accuracy of theoretical
models in predicting the $n$-$p$ capture cross section in the energy range of
interest for BBN.  Potentially, this could lead to a considerably smaller
uncertainty in $\Omega_Bh^2$ obtained from BBN.

We measured the analyzing power $\Sigma (90^{\circ})$ for the 
$^2$H$(\vec{\gamma},np)$ reaction with linearly polarized $\gamma$ rays at
$\theta = 90^{\circ}$ (lab) for seven energies between $E_{\gamma}$ = 2.39 and
4.05~MeV.  This energy range corresponds to $n$-$p$ c.m. energies of 165 keV to
1.83~MeV, i.e., the present experiment includes for the first time data in the
upper energy range of interest to BBN.  The analyzing power $\Sigma(\theta)$
is defined as
$$
\Sigma(\theta) = \frac{\sigma(\theta,\phi=0^{\circ}) - 
\sigma(\theta,\phi= 90^{\circ})}{\sigma(\theta,\phi=0^{\circ}) + 
\sigma(\theta,\phi= 90^{\circ})} \times  \frac{1}{f} =
\frac{b\sin^2 \theta}{a + b\sin^2 \theta} \times \frac{1}{f},
$$
where the differential cross section $\sigma(\theta,\phi)$ is given by
$$
\sigma(\theta,\phi) \sim a + b\sin^2\theta [1 + \cos 2\phi ].
$$
Here, $\theta$ is the polar angle, $\phi$ is the azimuthal angle, and $f$
is the degree of linear polarization of the incident $\gamma$-ray  beam.  The
quasi-monoenergetic and linearly polarized $\gamma$-ray beam was produced 
by Compton backscattering of relativistic electrons from 670 nm free-electron
laser (FEL) photons at the High-Intensity Gamma-ray Source (HIGS) located 
at the Duke University Free-Electron Laser Laboratory.  The electron energy
in the electron storage ring was varied between $E_e$ = 300 and 375~MeV to
generate $\gamma$-ray beams of energy between 2.39 and 4.05~MeV.  At a 
distance of
75 m from the electron-FEL-photon collision point the collimated 
$\gamma$-ray beam of 2.6 cm diameter struck a 4 cm diameter and 
6 cm long deuterated liquid scintillator (C$_6$D$_{12}$, Nuclear Enterprises
NE232) contained in a thin-walled glass container and viewed by a 
photomultiplier tube (PMT).
\begin{figure}
\includegraphics[width=3in]{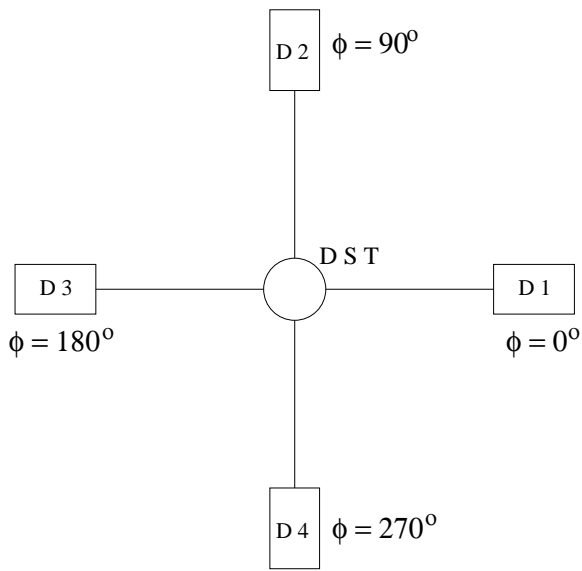}
\caption{Schematic of experimental setup.  The $\gamma$-ray beam is 
perpendicular into the page and the $\gamma$-ray polarization is nominal in
the $\phi=0$ plane.}
\label{fig1}
\end{figure}
The axis of the scintillator-PMT arrangement coincided with the axis of the 
incident $\gamma$-ray beam.  The average $\gamma$-ray flux at the location
of this deuterated scintillator target (DST) was $5 \times 10^5$
$\gamma$/s.
The $\gamma$-ray beam was monitored with a 140\% HPGe detector positioned
downstream of the experimental setup.  Aside from low-energy $\gamma$-ray 
sources
the ``natural'' $\gamma$-ray lines at $E_{\gamma}$ = 1461~keV ($^{40}$K) and
$E_{\gamma}$ = 2614.5~keV ($^{208}$Tl) served as convenient online calibration
sources throughout the course of the experiment.  The energy spread 
$\Delta E/E$ of the $\gamma$-ray beam varied between 2.3\% FWHM at $E_{\gamma}$
= 2.39~MeV to 2.9\% FWHM at $E_{\gamma}$ = 4.05~MeV.

\begin{figure}
\includegraphics[width=3in]{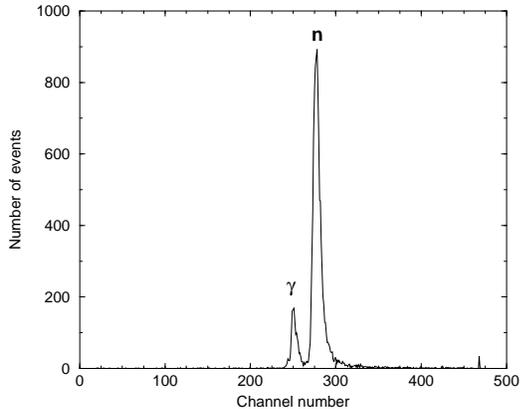}
\caption{Neutron time-of-flight spectrum between the deuterated scintillator
and a neutron detector for the reaction $\gamma$-$d \rightarrow n$-$p$ at 
$E_{\gamma}$ = 4.05~MeV.  Time increases from left to right.  The dominant 
peak is due to the neutrons of interest.  The small peak is due to $\gamma$
rays leaking through the PSD cut.  This leakage is smaller than 0.1\%.}
\label{fig2}
\end{figure}

\begin{figure}
\includegraphics[width=3in]{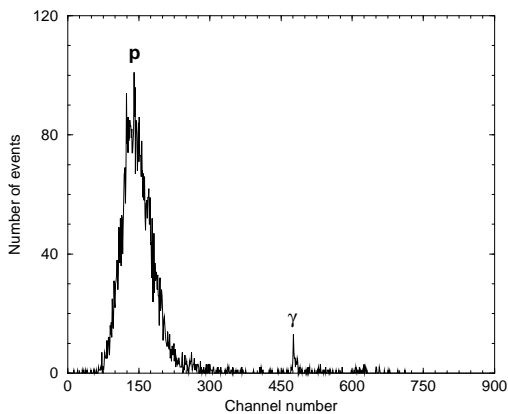}
\caption{Proton recoil energy spectrum in the deuterated scintillator (DST)
at $E_{\gamma}$ = 4.05~MeV.  The small peak near channel 450 is due to 
electrons generated via Compton scattering to the neutron detector.}
\label{fig3}
\end{figure}

The experimental setup is shown schematically in Fig.~\ref{fig1}.
Neutrons from the deuteron breakup reaction were detected by four Bicron 501A
liquid scintillator detectors, 2'' in diameter and 2'' in length, viewed by
a PMT.  We used four detectors rather than two to increase the efficiency of
our experimental setup.  Two neutron detectors were mounted at $\theta_{lab} =
90^{\circ}$ in the plane of the $\gamma$-ray polarization (nominally the
horizontal plane) on opposite sides of the incident $\gamma$-ray beam 
($\phi = 0^{\circ}$ and $180^{\circ}$).  The other two detectors were mounted 
at $\theta_{lab} = 90^{\circ}$ in the perpendicular plane ($\phi = 90^{\circ}$ 
and $270^{\circ}$). The center-to-center distance between the DST and the 
neutron detectors was 17 cm.  The protons from the deuteron breakup in the
DST gave the start signal for a neutron time-of-flight measurement between
the DST and the neutron detectors.  Neutron-gamma pulse-shape discrimination
(PSD) techniques were applied to distinguish the events of interest from the
overwhelming background produced in the neutron detectors by Compton scattering
from the DST.  Two-dimensional spectra of pulse height in the DST versus
neutron time-of-flight were created for the four neutron detectors used in the 
present experiment. Time-of-flight and proton recoil energy spectra are shown 
in Figs.~\ref{fig2} and \ref{fig3} for an incident gamma-ray energy of 
4.05~MeV. Our experimental techniques cannot be extended
to much lower $\gamma$-ray energies than already achieved in the present
experiment.  At $E_{\gamma}$ = 2.39~MeV both the proton and neutron energies 
were only 90 keV compared to the more comfortable value of about 900 keV at 
$E_{\gamma}$ = 4.05~MeV.  Liquid scintillator detectors are not commonly
employed to detect neutrons and protons at energies of less than 500~keV.
However, other types of detectors lack the fast timing characteristics and 
efficiencies that are crucial for obtaining $\gamma$-$d$ data in the energy 
range between $E_{\gamma}$ = 2.4 and 3~MeV.

In order to cancel instrumental asymmetries in our experimental setup, we 
either rotated the neutron detectors which were mounted on a ring centered and
positioned perpendicular to the $\gamma$-ray beam axis through $90^{\circ}$ 
(counter clockwise),
or we interchanged the detectors, i.e., the detectors in the horizontal 
plane were moved to the vertical plane and vice versa.  Within statistical
uncertainties either procedure gave consistent results for the asymmetry
$\epsilon$, which was calculated from the formula $\epsilon = (\alpha - 1)/
(\alpha + 1)$. 
For the rotation procedure we defined $\alpha$ as  
$\alpha_{1-2} = [(N_1^{HR}N_2^{HL})/(N_2^{VU}N_1^{VU})]^{1/2}$ for detector 
pair 1-2, and as 
$\alpha_{3-4} = [(N_3^{HL}N_4^{HR})/N_4^{VD}N_3^{VD})]^{1/2}$ 
for detector pair 3-4. Similarly, for the interchange procedure we have
$\alpha_{1-2} = [(N_1^{HR}N_2^{HR})/(N_2^{VU}N_1^{VU})]^{1/2}$ 
for detector pair 1-2, and 
$\alpha_{3-4} = [(N_3^{HL}N_4^{HL})/(N_4^{VD}N_3^{VD})]^{1/2}$ 
for detector pair 3-4. Here, $N_i^{HR}(N_i^{HL})$ refer to the neutron 
yields detected with detector $i$ positioned in the horizontal plane to the 
right (left) side of the incident $\gamma$-ray beam, and $N_i^{VU}(N_i^{VD})$ 
refer to the neutron yields detected with detector $i$ positioned in the 
vertical plane in the up (down) position. 

Based on the geometry of the undulator magnets used to produce the FEL
photons, the photon polarization should be linear and of magnitude 1.0.
Furthermore, the polarization plane should coincide with the horizontal 
plane in the
laboratory. Using the polarization dependent formulas for inverse Compton
scattering this should result in a linear $\gamma$-ray polarization in the
horizontal plane of $f = 1.0$ in the photon and electron 
energy range of interest for the present experiment.  However, the optical
cavity mirrors used to produce the FEL photons of 670 nm were optically active,
causing a rotation of the polarization plane of the FEL photons and 
consequently of the resulting $\gamma$-ray beam.  The outcoupled FEL light
was used to verify the linear polarization of 1.0 and to determine the tilt
angle of the polarization plane.  However, there is no guarantee that the tilt
angle of the polarization plane of the outcoupled light (i.e., the transmitted
light through the mirror opposite to the $\gamma$-ray beam direction) is in 
perfect agreement with the tilt angle of the FEL photon polarization inside
of the optical cavity where the electron-photon collision takes place.
Therefore, the tilt angle of the $\gamma$-ray polarization plane relative to 
the nominal horizontal plane was determined from the measured asymmetry
$\epsilon$ of the Compton scattered $\gamma$ rays by setting the PSD gate on 
the $\gamma$ rays in the neutron detectors and by selecting the appropriate 
pulse height gate (due to $\gamma$-ray scattering from electrons through
$90^{\circ}$) in the DST.  This asymmetry $\epsilon$ was determined 
simultaneously with the one for the breakup neutrons from the $\gamma$-$d$ 
reaction.  In order to extract the tilt angle from the measured $\gamma$-ray 
asymmetry data, the effective analyzing power of our apparatus for Compton
scattering from electrons was calculated via Monte-Carlo simulation using 
the Klein-Nishina formula. The polarization in multiple Compton scattering
was treated exactly.  The average tilt angle of the $\gamma$-ray 
polarization plane was found to be $(13.7 \pm 0.2)^{\circ}$ in upward 
direction relative to the horizontal laboratory plane.  This value is about
$2^{\circ}$ larger than the polarization tilt angle of the outcoupled FEL 
photons.

The neutron asymmetry data from the $\gamma$-$d \rightarrow n$-$p$ reaction
were corrected for finite geometry and multiple-scattering effects via 
extensive Monte-Carlo simulations of the experimental setup, using the tilt 
angle of the $\gamma$-ray polarization determined above and the $\gamma$-$d$
cross section and analyzing power calculations of Arenh{\"o}vel \cite{Are} 
which
are based on the Bonn nucleon-nucleon potential model \cite{Mac87}.  The use of
an active deuterium target makes our data practically insensitive to 
multiple $\gamma$-ray scattering (i.e., Compton scattering off electrons) 
in the DTS before the $\gamma$-$d \rightarrow
n$-$p$ reaction of interest is taking place.  The light output produced by
the recoil electrons generated in the Compton scattering process is 
considerably larger than the light output produced by the protons from
the $\gamma$-$d \rightarrow n$-$p$ reaction.  Therefore, multiple $\gamma$-ray
scattering can be eliminated efficiently by setting a tight gate on the proton 
pulse height of interest.  In contrast, multiple
scattering of the neutrons from the $\gamma$-$d$ reaction in the
DST has to be taken seriously.  Especially 
at the lowest $\gamma$-ray energies employed in the present experiment, our
constraint on the proton pulse height in the DST and our cut on the neutron
time-of-flight did not eliminate multiple scattering events completely due
to limitations of the detectors' energy and time resolutions.

\begin{figure}
\includegraphics[width=3in,clip]{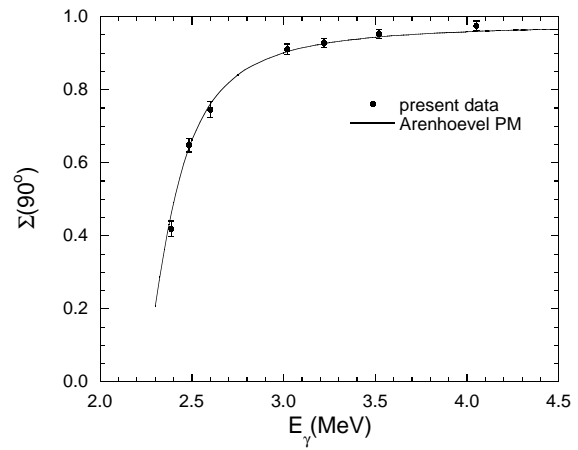}
\caption{Excitation function of the photon analyzing power $\Sigma$ for the
reaction $\gamma$-$d \rightarrow n$-$p$ at $\theta_{lab} = 90^{\circ}$ in
comparison to the theoretical prediction of Arenh{\"o}vel.}
\label{fig4}
\end{figure}

The data for the excitation function of the analyzing power $\Sigma 
(90^{\circ}$ lab) are shown in Fig.~\ref{fig4} and listed in Table~\ref{tab1}.
The error bars include statistical and systematic uncertainties added in 
quadrature.  At the higher energies the analyzing power is close to 1, i.e.,
the neutrons are emitted almost completely in the plane of the $\gamma$-ray
polarization (electric dipole radiation $E1$).  At energies below
$E_{\gamma}$ = 3~MeV, $\Sigma (90^{\circ})$ decreases rapidly, i.e., the
probability for neutrons to be emitted in the vertical plane (magnetic 
dipole radiation M1) increases with decreasing $\gamma$-ray energy.  The curve
shown in Fig.~\ref{fig4} is the prediction of Arenh{\"o}vel \cite{Are} using 
the coordinate-space version of the Bonn nucleon-nucleon potential model 
\cite{Mac87}.
The calculation includes meson-exchange, isobar, and relativistic effects.
Clearly, the model calculation is in very good agreement with the 
experimental data. Table~\ref{tab1} shows that the effective field theory
approach of Chen and Savage \cite{Che99} gives basically the same results as 
the potential model calculation of Arenh{\"o}vel.

\begin{widetext}
\begin{table}[h]
\caption{Measured photon analyzing power $\Sigma$ at $\theta_{lab} = 
90^{\circ}$ in comparison to theoretical predictions.}
\begin{tabular}{cccccc}
$E_{\gamma}$ (MeV) & $\theta_{c.m.}$ (deg) & $E^{n-p}_{c.m.}$ (keV) & 
$\Sigma$ &
$\Sigma_{\rm Arenhoevel}$ & $\Sigma_{\rm Chen{\&}Savage}$ \\
2.39 & 95.6 & 165 & $0.419 \pm 0.021$ & 0.461 & 0.464 \\
2.48 & 94.6 & 255 & $0.649 \pm 0.019$ & 0.624& 0.631 \\
2.60 & 94.0 & 375 & $0.745 \pm 0.022$ & 0.760 & 0.757 \\
3.02 & 93.2 & 795 & $0.911 \pm 0.014$ & 0.902 & 0.901 \\
3.22 & 93.0 & 995 & $0.928 \pm 0.012$ & 0.925 & 0.923 \\
3.52 & 92.9 & 1295 & $0.953 \pm 0.012$ & 0.944 & 0.942 \\
4.05 & 92.8 & 1825 & $0.975 \pm 0.013$ & 0.959 & 0.958 \\
\end{tabular}
\label{tab1}
\end{table}
\end{widetext}

\begin{figure}
\includegraphics[width=3in]{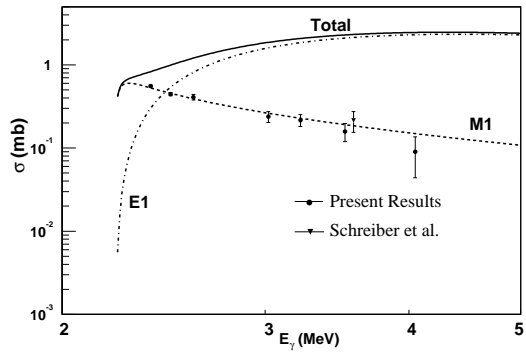}
\caption{Data for the M1 contribution to the $\gamma$-$d$ total cross section 
in comparison to the theoretical prediction of Chen and Savage (dashed curve).
The dashed-dotted curve represents the E1 contribution and the solid curve in
the total $\gamma$-$d$ cross section from Chen and Savage.  Note the 
logarithmic scales for both $\sigma$ and $E_{\gamma}$.}
\label{fig5}
\end{figure}

As shown in detail by Schreiber {\em et al}. \cite{Sch00} the $\gamma$-$d$ 
analyzing power data $\Sigma(\theta)$ at low energies can be used to determine
the relative M1 and E1 strengths of the $\gamma$-$d$ cross section. Based
on the present $\Sigma$ ($90^{\circ}$ lab) data, Table~\ref{tab2} gives the
calculated M1 contribution to the $\gamma$-$d$ cross section in comparison
to the effective-field theory calculations of Chen and Savage, Rupak 
\cite{Rup00}, and the
nucleon-nucleon potential-model calculation of Arenh{\"o}vel. In the energy
range most important for BBN our results
are in excellent agreement with the theoretical predictions,
especially with the calculations of Arenh{\"o}vel.  Fig.~\ref{fig5} shows
the calculated total $\gamma$-$d$ cross section of Chen and Savage
as well as the 
associated M1 and E1 contributions in comparison to the M1 contribution 
determined in the present work (dots) and in the earlier work of Schreiber 
{\em et al}. at $E_{\gamma}$ = 3.58~MeV (triangle).

\begin{table}
\caption{M1 (s-wave) contribution S to the total $\gamma$-$d$
cross section obtained from the present $\Sigma(\theta)$ data in comparison
to the predictions of Arenh{\"o}vel, Chen and Savage, and Rupak.} 
\begin{tabular}{c|cccc}
$E_{\gamma}$ & \multicolumn{4}{c}{$S$}\\
(MeV) & this experiment & Arenh{\"o}vel & Chen {\&} Savage & Rupak\\
\hline
2.39 & $0.675 \pm 0.019$ & 0.662 & 0.622 & 0.627\\
2.48 & $0.448 \pm 0.020$ & 0.468 & 0.459 & 0.458\\
2.60 & $0.339 \pm 0.026$ & 0.328 & 0.320 & 0.317\\
3.02 & $0.128 \pm 0.019$ & 0.141 & 0.139 & 0.135\\
3.22 & $0.104 \pm 0.017$ & 0.108 & 0.109 & 0.104\\
3.52 & $0.069 \pm 0.017$ & 0.080 & 0.083 & 0.079\\
4.05 & $0.037 \pm 0.019$ & 0.057 & 0.061 & 0.056\\
\end{tabular}
\label{tab2}
\end{table}

In summary, the first experimental test of theoretical models used to 
calculate the $n$-$p$ capture cross section in the energy range of importance
to BBN reveals almost perfect agreement with experimental information 
derived from analyzing power data for the reverse reaction $\gamma$-$d 
\rightarrow n$-$p$.  This observation lends substantial credibility to
the theoretical models also in the presently not tested $\gamma$-ray
energy range from 2.25 to 2.38~MeV, i.e., for $n$-$p$ c.m. energies between
25 and 155 keV.  Work is planned to reduce the uncertainty of the described
measurements from its present 3\% uncertainty at 2.39~MeV in determining the
dominant M1 contribution to the $\gamma$-$d$ cross section to 
1.5\% and to extend the measurements to $n$-$p$ c.m. energies as low as 
25~keV.

We conclude that the $\pm 5\%$ uncertainty used in \cite{Bur99} is a very 
conservative estimate for the uncertainty of modern theoretical approaches
available for calculating the $n$-$p$ capture cross section in the energy range
relevant to BBN.  
The uncertainty quoted in Ref.~\cite{Bur99} can be reduced by at least 
20\%. The
planned improvements of our measurements are expected to provide an even more
accurate test of the calculated $n$-$p$ capture cross section.  Therefore, this
cross section will play a small role in the overall uncertainty of the 
baryon density in the early universe as determined in Ref.~\cite{Bur99}.  
The BBN approach compares favorably with the very recent CMB 
based method of determining $\Omega_Bh^2$ from the WMAP data \cite{Ben,Spe}.

\begin{acknowledgments}
This work was supported in part by the US Department of Energy, Office of
High-Energy and Nuclear Physics, under grant No. DE-FG02-97ER41033 and
DE-FG02-97ER41042.  N.G.C. acknowledges
partial support from the US National Science Foundation REU Program 
PHY-9912252. The authors would like to thank M.W. Ahmed and J.H.
Esterline for their contributions to the present work.  
\end{acknowledgments}

\bibliography{bbn_bib}

\end{document}